\newcommand{\CLASSINPUTtoptextmargin}{0.75in}%
\newcommand{\CLASSINPUTbottomtextmargin}{1in}%
\newcommand{\CLASSINPUTinnersidemargin}{0.63in}%
\newcommand{\CLASSINPUToutersidemargin}{0.63in}%
\documentclass[conference,10pt,letterpaper]{IEEEtran}%
\usepackage{amsmath}
\usepackage{graphicx}
\usepackage{multirow}
\usepackage[none]{hyphenat}
\usepackage{float}
\usepackage{subfig}
\usepackage{dblfloatfix}
\usepackage{eso-pic}
\usepackage{t1enc}
\usepackage{times}

\makeatletter

\def\@IMSauthorblockNAMEstyle{\normalfont\IMSauthorsize}
\def\@IMSauthorblockAFFILstyle{\normalfont\IMSaffilsize}
\def\@IMSauthorblockEMAILstyle{\normalfont\IMSaffilsize}
\def\IMSauthorblockNAME#1{%
\relax\@IMSauthorblockNAMEstyle%
#1%
}%
\def\IMSauthorblockAFFIL#1{%
\relax\@IMSauthorblockAFFILstyle%
\vskip\@IEEEauthorblockAtopspace
#1%
}%
\def\IMSauthorblockEMAIL#1{%
\relax\@IMSauthorblockEMAILstyle%
\vskip\@IEEEauthorblockAtopspace
#1%
}%
\newcommand{\IMSauthor}[1]{%
\ifIsBlindReviewVersion%
\author{\phantom{\parbox{\textwidth}{\center\relax#1}}}%
\else%
\author{\parbox{\textwidth}{\center\relax#1}}%
\fi%
}%
\newif\ifIsBlindReviewVersion

\def\IMSthispaperforfinalpublication{\IsBlindReviewVersionfalse}
\def\@maketitle{\newpage
\bgroup\par\addvspace{0.5\baselineskip}\centering%
\ifCLASSOPTIONtechnote
   {\bfseries\large\@IEEEcompsoconly{\sffamily}\@title\par}\vskip 1.3em{\lineskip .5em\@IEEEcompsoconly{\sffamily}\@author
   \@IEEEspecialpapernotice\par{\@IEEEcompsoconly{\vskip 1.5em\relax
   \@IEEEtitleabstractindextextbox{\@IEEEtitleabstractindextext}\par
   \hfill\@IEEEcompsocdiamondline\hfill\hbox{}\par}}}\relax
\else
   \vskip0.2em{\IMStitlesize\ifCLASSOPTIONtransmag\bfseries\LARGE\fi\@IEEEcompsoconly{\sffamily}\@IEEEcompsocconfonly{\normalfont\normalsize\vskip 2\@IEEEnormalsizeunitybaselineskip
   \bfseries\Large}\@title\par}\vskip1.0em\par
   \ifCLASSOPTIONconference%
      {\@IEEEspecialpapernotice\mbox{}\vskip\@IEEEauthorblockconfadjspace%
       \mbox{}\hfill\begin{@IEEEauthorhalign}\@author\end{@IEEEauthorhalign}\hfill\mbox{}\par}\relax
   \else
      \ifCLASSOPTIONpeerreviewca
         {\@IEEEcompsoconly{\sffamily}\@IEEEspecialpapernotice\mbox{}\vskip\@IEEEauthorblockconfadjspace%
          \mbox{}\hfill\begin{@IEEEauthorhalign}\@author\end{@IEEEauthorhalign}\hfill\mbox{}\par
          {\@IEEEcompsoconly{\vskip 1.5em\relax
           \@IEEEtitleabstractindextextbox{\@IEEEtitleabstractindextext}\par\hfill
           \@IEEEcompsocdiamondline\hfill\hbox{}\par}}}\relax
      \else
         \ifCLASSOPTIONtransmag
           {\@IEEEspecialpapernotice\mbox{}\vskip\@IEEEauthorblockconfadjspace%
            \mbox{}\hfill\begin{@IEEEauthorhalign}\@author\end{@IEEEauthorhalign}\hfill\mbox{}\par
           {\vspace{0.5\baselineskip}\relax\@IEEEtitleabstractindextextbox{\@IEEEtitleabstractindextext}\vspace{-1\baselineskip}\par}}\relax
         \else
           {\lineskip.5em\@IEEEcompsoconly{\sffamily}\sublargesize\@author\@IEEEspecialpapernotice\par
           {\@IEEEcompsoconly{\vskip 1.5em\relax
            \@IEEEtitleabstractindextextbox{\@IEEEtitleabstractindextext}\par\hfill
            \@IEEEcompsocdiamondline\hfill\hbox{}\par}}}\relax
         \fi
      \fi
   \fi
\fi\par\addvspace{0.0\baselineskip}\egroup}

\def\IMStitlesize{\@setfontsize{\IMStitlesize}{18}{21pt}}
\def\IMSauthorsize{\@setfontsize{\IMSauthorsize}{12}{13pt}}
\def\IMSaffilsize{\@setfontsize{\IMSaffilsize}{12}{13pt}}
\def\IMScaptionsize{\@setfontsize{\IMScaptionsize}{8}{9pt}}
\def\IMSbibsize{\@setfontsize{\IMSbibsize}{8}{9pt}}

\def\@IEEEauthorblockNstyle{\IMSauthorsize\@IEEEcompsocnotconfonly{\sffamily}\@IEEEcompsocconfonly{\large}}
\def\@IEEEauthorblockAstyle{\IMSaffilsize\@IEEEcompsocnotconfonly{\sffamily}\@IEEEcompsocconfonly{\itshape}\@IEEEcompsocconfonly{\large}}
\def\@IEEEauthordefaulttextstyle{\IMSauthorsize\@IEEEcompsocnotconfonly{\sffamily}\sublargesize}

\def\thebibliography#1{\section*{\refname}%
    \addcontentsline{toc}{section}{\refname}%
    \IMSbibsize\@IEEEcompsocconfonly{\small}\vskip 0.3\baselineskip plus 0.1\baselineskip minus 0.1\baselineskip
    \list{\@biblabel{\@arabic\c@enumiv}}%
    {\settowidth\labelwidth{\@biblabel{#1}}%
    \leftmargin\labelwidth
    \advance\leftmargin\labelsep\relax
    \itemsep \IEEEbibitemsep\relax
    \usecounter{enumiv}%
    \let\p@enumiv\@empty
    \renewcommand\theenumiv{\@arabic\c@enumiv}}%
    \let\@IEEElatexbibitem\bibitem%
    \def\bibitem{\@IEEEbibitemprefix\@IEEElatexbibitem}%
\def\newblock{\hskip .11em plus .33em minus .07em}%
\ifCLASSOPTIONtechnote\sloppy\clubpenalty4000\widowpenalty4000\interlinepenalty100%
\else\sloppy\clubpenalty4000\widowpenalty4000\interlinepenalty500\fi%
    \sfcode`\.=1000\relax}

\long\def\@makecaption#1#2{%
\ifx\@captype\@IEEEtablestring%
\par\@IEEEtabletopskipstrut
\else
\@IEEEfigurecaptionsepspace
\fi
\setbox\@tempboxa\hbox{\normalfont\IMScaptionsize {#1.}\nobreakspace\nobreakspace #2}%
\ifdim \wd\@tempboxa >\hsize%
\setbox\@tempboxa\hbox{\normalfont\IMScaptionsize {#1.}\nobreakspace\nobreakspace}%
\parbox[t]{\hsize}{\normalfont\IMScaptionsize\noindent\unhbox\@tempboxa#2}%
\else
\ifCLASSOPTIONconference \hbox to\hsize{\normalfont\IMScaptionsize\hfil\box\@tempboxa\hfil}%
\else \hbox to\hsize{\normalfont\IMScaptionsize\box\@tempboxa\hfil}%
\fi\fi
\ifx\@captype\@IEEEtablestring%
\@IEEEtablecaptionsepspace
\else
\fi}

\newlength\tablecaptiontotableskip
\newlength\figuretocaptionskip
\def\@IEEEfigurecaptionsepspace{\vskip\figuretocaptionskip\relax}%
\def\@IEEEtablecaptionsepspace{\vskip\tablecaptiontotableskip\relax}%

\def\abstract{\normalfont%
\@IEEEabskeysecsize\bfseries\textit{\abstractname}\,\bfseries\textit{---}\,%
\@IEEEgobbleleadPARNLSP}%

\def\IEEEkeywords{\normalfont%
\@IEEEabskeysecsize\bfseries\textit{\IEEEkeywordsname}\,\bfseries\textit{---}\,%
\@IEEEgobbleleadPARNLSP}%
\def\endIEEEkeywords{\relax\vspace{0.67ex}%
\par\if@twocolumn\else\endquotation\fi%
\normalsize\normalfont}%

\def\@IEEEauthorblockNtopspace{0ex}
\def\@IEEEauthorblockAtopspace{1mm}
\def\tablename{Table}
\def\thetable{\arabic{table}}
\def\IEEEkeywordsname{Keywords}
\def\subsubsection{\@startsection{subsubsection}{3}{\z@}{1.5ex plus 1.5ex minus 0.5ex}%
{0.7ex plus .5ex minus 0ex}{\normalfont\normalsize\itshape}}%
\def\@seccntformat#1{\csname the#1dis\endcsname\relax}
\def\thesectiondis{\thesection.\hskip 0.5em}
\def\thesubsectiondis{{\hbox to\parindent{\Alph{subsection}.}}}
\def\thesubsubsectiondis{{\hbox to \parindent{\arabic{subsubsection})}}}
\def\theparagraphdis{{\hbox to \parindent{\alph{paragraph})}}}
\IEEEilabelindentA \parindent
\IEEEilabelindent \IEEEilabelindentA
\IEEEelabelindent \parindent
\IEEEdlabelindent \parindent
\IEEElabelindent \parindent

\newlength\@IMSparindent
\newcommand\IMSdisplayacksection[1]{%
\ifIsBlindReviewVersion%
\noindent\phantom{\parbox[t]{\columnwidth}{\normalbaselines\setlength{\parindent}{\@IMSparindent}{#1}\strut}}
\else%
\noindent\parbox[t]{\columnwidth}{\normalbaselines\setlength{\parindent}{\@IMSparindent}{#1}\strut}%
\fi%
}%

\makeatother

\begin{document}
\raggedbottom
%
%
%
\title{Deep-Learning-Based Pixelated Microwave Filter Design and Characterization using Electro-Optical Electric-Field Measurements}


%
%
%
\IMSthispaperforfinalpublication
\IMSauthor{%
\IMSauthorblockNAME{
Han Zhou\textsuperscript{\protect\#}, Richard Bannister\textsuperscript{\protect\$}, Caspar Pierce\textsuperscript{\protect\$}, Haojie Chang\textsuperscript{\protect\#}, David Widén\textsuperscript{\protect\#}, Ludvig Fornstedt\textsuperscript{\protect\#},\\ 
Gabriel~Melin\textsuperscript{\protect\#}, Alexander Bohlin\textsuperscript{\protect\#}, Pontus Lindeberg Fredriksson\textsuperscript{\protect\#}, \\ Dilbagh Singh\textsuperscript{*},
Christian Fager\textsuperscript{\protect\#}, Koen~Buisman\textsuperscript{\protect\$}
}
\\%
\IMSauthorblockAFFIL{
\textsuperscript{\protect\#}Chalmers University of Technology, Gothenburg, Sweden
}
\IMSauthorblockAFFIL{
\textsuperscript{\protect\$}Advanced Technology Institute, University of Surrey, Guildford, UK}
\\%
\IMSauthorblockAFFIL{
\textsuperscript{*}National Physical Laboratory, Teddington, UK.}
\\%

\IMSauthorblockEMAIL{
han.zhou@ieee.org, k.buisman@surrey.ac.uk\\
}
}
%
\AddToShipoutPictureFG*{%
  \AtPageUpperLeft{%
    \hspace{0.63in}%
    \raisebox{-0.32in}{%
      \parbox{\dimexpr\paperwidth-1.26in\relax}{%
        \centering\footnotesize\itshape
        This is the author-accepted version of a paper accepted for presentation at the
        2026 IEEE/MTT-S International Microwave Symposium (IMS 2026).
        \copyright~2026 IEEE. The published version is available at https://doi.org/10.1109/IMSRFSA70221.2026.11624364
      }%
    }%
  }%
}

\maketitle
%
%
%
\begin{abstract}
Traditional microwave filter design typically relies on iterative parameter tuning and predefined topologies, which limits design space and increases development time. This study uses a deep learning approach combining convolutional neural networks with genetic algorithms to automate pixelated microwave filter synthesis. To validate the approach experimentally, both S-parameter and spatial electric-field measurements were analyzed. The synthesized low-pass filter demonstrated excellent agreement between simulated and measured performance, achieving a 7~GHz passband with over 20~dB suppression beyond 9.5~GHz. Electro-optical measurements, for the first time, revealed electric field patterns that resemble coupled transmission-lines or stub structures, providing insight into the emergent characteristics of AI-generated designs.

\end{abstract}
%
%

\section{Introduction}
The rapid growth of data throughput has driven wireless communication systems toward higher frequencies and greater integration. This shift has heightened the importance of designing complex electromagnetic (EM) layout structures. Traditional design methods for microwave circuits primarily rely on pre-selected EM topologies informed by prior knowledge of physical effects, employing either lumped or distributed elements. While effective, these approaches demand meticulous tailoring of parameterized layouts, making the process time-intensive and limiting the design space.

In recent years, a method has been introduced that pixelates the geometry of EM layout structures into binary matrices, where each pixel represents the presence or absence of metal. When combined with deep learning techniques, this approach shows significant potential to expand the design space and surpass traditional design methodologies. This method has been successfully applied in various domains, including integrated photonics~\cite{Ma2020DeepLF}, metalens antennas~\cite{DL_Meta}, and millimeter-wave passive and integrated circuits~\cite{PA_DL_NC}. These non-intuitive pixelated geometries operate through unobvious EM coupling. Electric field measurements can offer insight into the operation and emergent characteristics of such pixelated designs.

Existing studies primarily validate artificial intelligence (AI) designed circuits through simulated or measured S-parameters~\cite{DL_PA, AI_HZ, AI_F-1_HZ, AI_Chenhao, AIDPA_HZ}, which quantify port responses but do not give insight into their EM behavior. Different from human-engineered layouts, AI synthesized pixelated designs may exploit unobvious EM interactions.
To bridge this gap, electro-optical (EO) field measurements provide a non-intrusive method to spatially resolve electric fields across broad bandwidths~\cite{Sarabandi2017}. Unlike conventional near-field probes, EO techniques map vectorial electric fields without perturbing device operation. Despite their success in characterizing traditional passive~\cite{EO_KIT, Kamal} and active~\cite{Peter} components, EO methods remain unexplored in analyzing AI-driven EM designs, particularly in understanding how non-traditional geometries manipulate electric fields.

This research focuses on this area by presenting EO measurements and their interpretation for an AI-designed low-pass filter, thereby gaining understanding and learning from these AI-designed circuits.

\begin{figure} [t]
    \centering       \includegraphics[width=0.95\columnwidth]{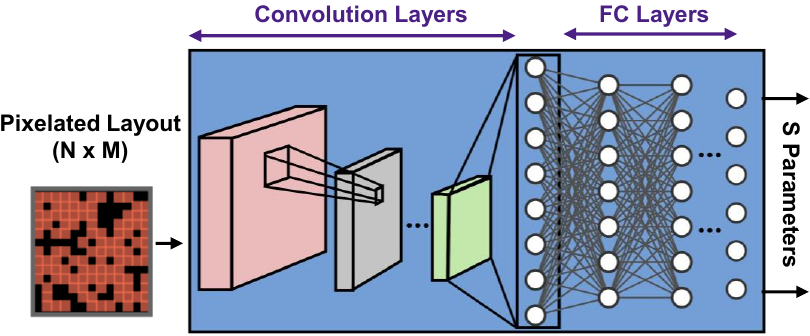}
    \caption{The trained deep CNN architecture with a binary matrix input representing the pixelated EM layout and S-parameters as output. The forward emulator, derived from the trained CNN-based surrogate model, enables the inverse synthesis of EM structures to achieve desired S-parameters.}
    \label{fig.1}
\end{figure}

\section{Method}
\subsection{Deep-Learning-Based Filter Design}
The employed deep learning method uses a planar circuit layout discretized into a grid, where each cell is binary: "1" for metal and "0" for non-metal. The layout structure is discretized into a $13\mathrm{\times}13$ grid of pixels, with a pixel dimension of $0.9\mathrm{\times}0.9~\mathrm{mm}$. This configuration balances pixel coupling effects and training resources. Reliable electrical contact is ensured through a 10 \% metal overlap. We generate 100k pixelated layouts in Keysight Advanced Design System Momentum using a Python script and expand the dataset to 400k through rotation and mirroring. We allocate 90 \% of the dataset for training and 10 \% of the dataset for validation.  The metal density follows a normal distribution (mean: 50~\%, dev: 15 \%), with metal coverage constrained to 10 -- 80 \%.
The large dataset of circuits and their S-parameters trains a deep convolutional neural network (CNN). This CNN generates candidate solutions evaluated in a genetic algorithm ~\cite{GA} to meet target S-parameters. We train the CNN on an Nvidia 4070 Super, completing 600 epochs in $8$ hours. The 600 epochs ensure sufficient exposure to diverse training input for robust performance. The trained CNN achieves a mean absolute validation error of 3.2\%.

\begin{figure} [t]
    \centering       
    \includegraphics[width=0.95\columnwidth]{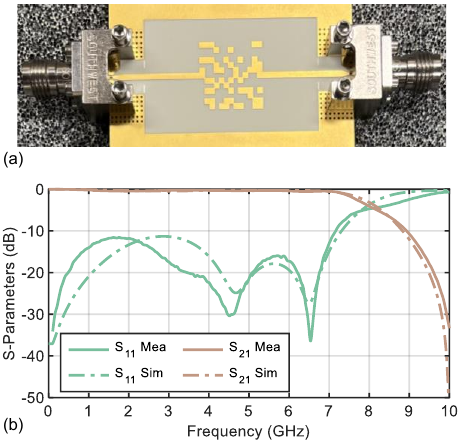}
    \caption{ (a) The deep-learning-based realisations of a low-pass filter along with (b) its simulated and measured S-parameters over the design frequency.}
    \label{fig.2}
\end{figure}

As illustrated in Fig.~\ref{fig.1}, the CNN is designed to process a $13\mathrm{\times}13$ binary matrix as input and predict the corresponding S-parameters. The output consists of the complex parts of $S_{11}$, $S_{12}$ and $S_{22}$ at $19$ frequency points between $1$ and $10~\mathrm{GHz}$, resulting in an output size of $114$. The CNN architecture comprises $14$ convolutional layers, each with $64$ filters, with filter sizes decreasing from $6\mathrm{\times}6$ to $3\mathrm{\times}3$. These layers are organized into seven pairs, where each pair integrates residual connections to both the preceding and earlier layers, enhancing stability and accelerating convergence \cite{He_2016_CVPR}. The extracted features are then passed through $4$ fully connected (FC) layers, each containing $2048$ neurons. To improve nonlinear modeling capabilities, Leaky ReLU activation functions are employed, allowing small gradients for negative inputs, while batch normalization ensures stable training by normalizing activations at each layer ~\cite{ReLU}. The extracted features are subsequently flattened before being passed through the fully connected layers. To reduce overfitting and improve generalization, dropout layers with a dropout rate of $30$ \% are integrated into the fully connected layers \cite{Dropout}. 

\subsection{Electro-Optical Electric-Field Measurements}

The EO field scanner used can measure electric fields outside radio-frequency (RF) components, up to 40 GHz. However, due to limitations in the power amplifier driving the device-under-test (DUT), the measurable frequency range is restricted to 200 MHz -- 10 GHz. To determine the vectorized field components, two EO probes are employed: one for measuring the normal component of the electric field and another for the tangential components. 

\begin{figure}[t!]
    \centering    
    \includegraphics[width=\columnwidth]{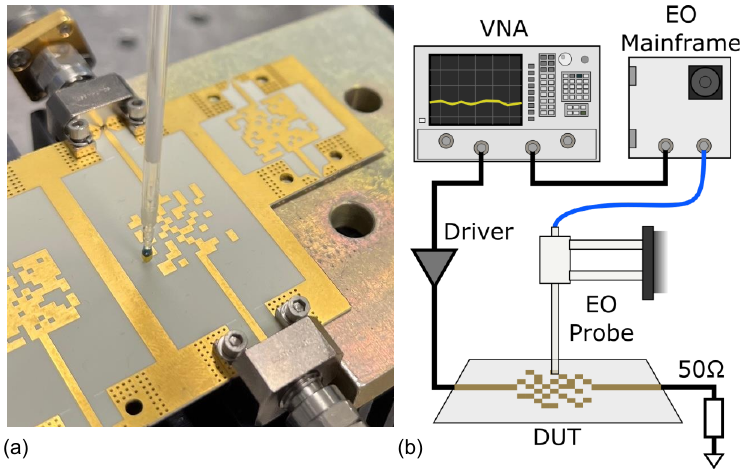}
    \caption{(a) EO field probe scanning over the low-pass filter. (b) Schematic of the measurement system.}
    \label{fig:image_EO_scan}
\end{figure}

\begin{figure*} [t!]
  \centering
  \includegraphics[width=\textwidth]{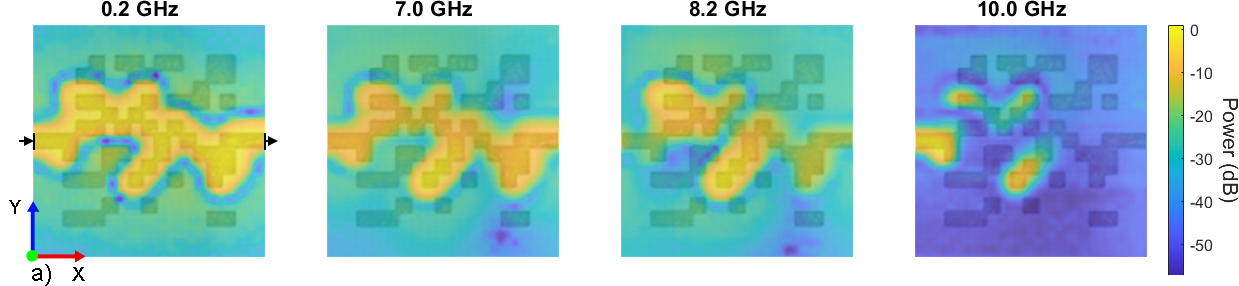}
  \includegraphics[width=\textwidth]{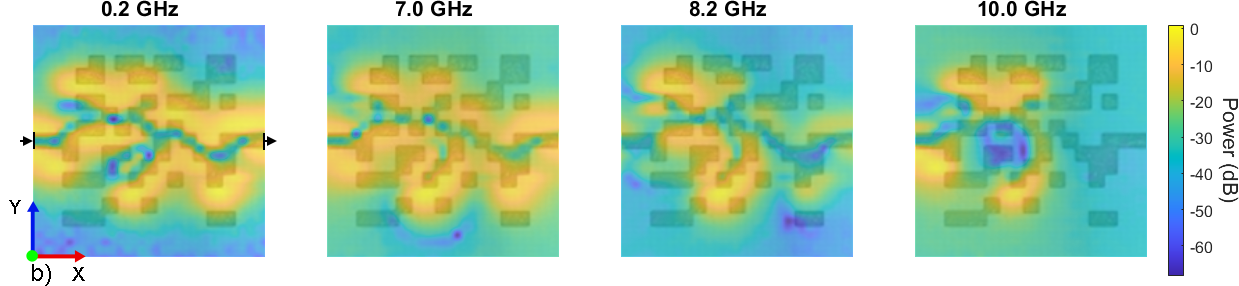}
  \includegraphics[width=\textwidth]{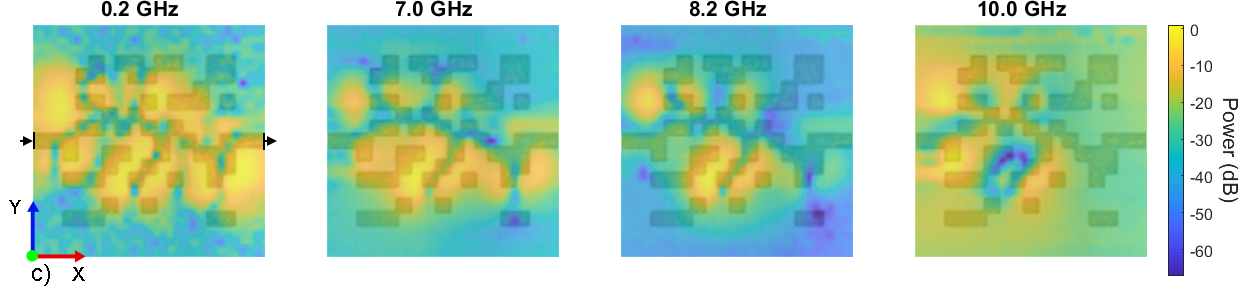}
  \caption{Normalized measured (a) $E_Z$, (b) $E_Y$ and (c) $E_X$ from $S^{EO}_{21}$ in dB mapped over a photo (dimension: $15.3\mathrm{\times}15.3$ mm) of the filter for four frequencies. The input and output are indicated with black arrows. The data is normalized at each frequency to maximize contrast. The orientation of x and y is indicated in the left figures.}
  \label{fig_E_freq}
\end{figure*}

The measurement principle utilizes the Pockels effect: when light passes through a Pockels crystal, its polarization changes with the applied electric field strength. This change enables measurement by analyzing light reflected at the probe's end, traveling back through the fiber, which is then converted to amplitude within the EO mainframe (Emagtech Neoscan)~\cite{Wu1998}. The resulting signal is measured using an external instrument, here a vector network analyzer (VNA). The probe consists predominantly of glass, thus allowing non-intrusive measurement, and is connected by single-mode fiber to the EO mainframe. The mainframe contains the required laser, polarizers and photodiodes~\cite{Sarabandi2017}. The laser operates at 1550~nm with a spot size of approximately 8 micrometers within the Pockels crystal, which dictates the system's resolution. To scan across the DUT, the EO probe is moved using a controlled xy linear stage with micron-scale position capability (Fig. \ref{fig:image_EO_scan}(a)). The EO field scanner is mounted on an optical table with active vibration suppression.

In the measurement setup one port of the VNA is connected to a driver amplifier that feeds into the DUT, whose output is terminated with 50~$\Omega$. The second port of the VNA connects to the EO mainframe and measures the resulting RF signal, created by the electric field over the DUT (Fig.~\ref{fig:image_EO_scan}(b)). By using $S^{EO}_{21}$ any source power variations during the measurement are mitigated. Where $^{EO}$ indicates this measurand is obtained using the measurement system in Fig. \ref{fig:image_EO_scan}(b).

\section{Measurement results and discussion}

\subsection{Filter RF Performance}
The focus of our study is a low-pass filter. The filter was implemented on a $20$-mil Rogers $4350$B substrate, as shown in Fig.~\ref{fig.2}, along with its simulated and measured S-parameters ($S_{11}$ and $S_{21}$). The simulated and measured results demonstrate excellent agreement. The low-pass filter exhibits a passband up to $7.0~\mathrm{GHz}$ (insertion loss \textless ~$0.36~\mathrm{dB}$). The stopband suppression exceeds $20~\mathrm{dB}$ beyond $9.5~\mathrm{GHz}$.

\subsection{Electric-Field Results and Discussion}
Electric field measurements were made using a normal ($E_Z$) and a tangential ($E_X$ and $E_Y$) EO field probe.  Weak RF signals necessitate an IF bandwidth of 1 Hz to provide good contrast. The measurement step size of 0.45 mm was half the pixel size (0.9 mm). Fig.~\ref{fig_E_freq} shows the results for $E_Z$, $E_Y$, and $E_X$ where the normalized amplitude of $S^{EO}_{21}$ in dB is given as color map, while the background depicts the spatial photo measuring $15.3\mathrm{\times}15.3$ mm. The measured electric field (Fig. \ref{fig_E_freq}) can be divided into three frequency regions: passband (0.2, 7 GHz), 3-dB transition (8.2 GHz), and stopband (10~GHz). 

In the passband region, a continuous path from input to output is evident in $E_Z$, confirming quasi-TEM transmission-line behavior. All pixels along this path are connected, with some connecting at pixel corners. A gap, consisting of a single pixel empty area, surrounds this path, while additional metal-filled pixels extend beyond it. This gap and its surrounding pixels may enable slow wave propagation via capacitive coupling to adjacent non-connected pixels. 

As seen in Fig. \ref{fig_E_freq}(b) $E_Y$ exhibits a distinct minimum directly above the metal path and maxima in the regions flanking it. This field distribution reflects a quasi-TEM mode in a stripline transmission-line, characterized by a dominant vertical field ($E_Z$) over the conductor with strong fringe fields confined to the adjacent region. The reduced amplitude of $E_Z$ in the gap, combined with the amplitude of $E_Y$ which has its minimum above the continuous path and maximum alongside it, mimics the  field above a stripline transmission-line. 
The strong reduction in $E_Y$ and $E_X$ outside the isolated pixels outside of the gap further supports the role of pixel derived capacitive coupling in shaping the field. Outside the structures, the electric field is weak and relatively uniform, with lower frequencies showing a more uniform field distribution.

As the filter approaches the 3-dB frequency, standing wave patterns emerge within the structure. These patterns show alternating minima and maxima in $E_Z$, with the field becoming more contained towards the input of the filter. This behavior is indicative of resonance likely arising from capacitive coupling and metal inductance, and seems similar to a coupled transmission-line filter, likely due to the pixel placement allowed during the AI-algorithm, resulting in diagonal stub-like configurations. 
Given the filter's small physical dimensions of $0.3\mathrm{\times}0.3$ $\lambda$ at the 3-dB frequency, the observed scale of field variations, particularly the distance between minima and maxima, aligns with the compressed wavelength characteristic of slow-wave propagation, which is induced by the quasi-periodic capacitive loading and indicates a reduced phase velocity.

Beyond the 3-dB frequency the standing wave pattern intensifies, shifting further towards the input of the device, resulting in less power reaching the output. Notably, the $E_Y$ and $E_X$ field minima/maxima exhibit partial inversion compared to the 3-dB transition region, a phenomenon not observed in $E_Z$. This inversion is likely due to impedance discontinuity, mismatch, and resonant coupling in the stopband, causing a modal transition, created by the structure’s quasi-periodicity and pixel coupling. The high stopband attenuation results from resonance effects and the structure’s inability to support the required guided modes, causing strong reflections and destructive interference.

The emergence of structures resembling transmission-lines without predefined geometries highlights the algorithm’s ability to autonomously generate filter configurations. This contrasts with \cite{Zou2025} which starts with well-defined transmission-lines and discretizes them into pixelated designs.

\section{Conclusion}
Deep learning techniques can significantly expand the EM design space by synthesizing complex structures from pixelated binary matrices. Here, we combined AI-driven design with experimental EO electric field measurements to advance understanding of these novel RF circuits. A CNN, trained on randomized layouts and their EM-simulated S-parameters, guided the GA optimization of a low-pass filter. EO field measurements provided spatial insights into electric field distributions, revealing emergent phenomena such as transmission-line like RF paths in the passband, slow-wave propagation, standing waves at 3-dB frequencies, and energy confinement near the input in the stopband. These observations somewhat correspond to behaviors seen in conventional coupled transmission-line filters. The EO measurements highlight the frequency-dependent electric field behavior of the AI‑generated designs, offering a method to interpret their EM behavior. Such advancements enable AI in EM design, facilitating rapid innovation for next-generation wireless systems.

\section*{Acknowledgment}
This research was supported by Swedish Innovation Agency (VINNOVA), Sivers Semiconductors, and Chalmers University of Technology under Grant 2022-00863.

\newcommand{\IMSacktext}{%
Authors wish to acknowledge...
}



\bibliographystyle{IEEEtran}

\bibliography{IEEEabrv,IEEEexample}

\end{document}